\documentstyle[emulateapj]{article}
\tighten
\newcommand{\etal}{{\rm et al.~}}
\newcommand{\Mpc}{$h^{-1}$~{\rm Mpc}}

\def\apj{ApJ\ }
\def\apjl{ApJL\ } 
\def\apjs{ApJS\ }

\begin{document}

\title{Optical and X-ray clusters  as tracers of the supercluster-void
network. I Superclusters of Abell and X-ray clusters}

\author{M. Einasto\altaffilmark{1},
J. Einasto \altaffilmark{1},
E. Tago\altaffilmark{1},
V. M\"uller  \altaffilmark{2} \&  
H. Andernach\altaffilmark{3}}
\altaffiltext{1}{Tartu Observatory, EE-61602 T\~oravere, Estonia}
\altaffiltext{2}{Astrophysical Institute Potsdam, An der Sternwarte 16,
       D-14482 Potsdam, Germany}
\altaffiltext{3} {Depto.  de Astronom\'\i a, Univ.\ Guanajuato,
            Apdo.\ Postal 144, Guanajuato, C.P.\ 36000, GTO, Mexico}

\begin{abstract}

We study the distribution of X-ray selected clusters of galaxies with
respect to superclusters determined by Abell clusters of galaxies and
show that the distribution of X-ray clusters follows the
supercluster-void network determined by Abell clusters.  We find that
in this network X-ray clusters are more strongly clustered than other
clusters: the fraction of X-ray clusters is higher in rich
superclusters, and the fraction of isolated X-ray clusters is lower
than the fraction of isolated Abell clusters.  There is no clear
correlation between X-ray luminosity of clusters and their host
supercluster richness.  Poor, non-Abell X-ray clusters follow the
supercluster-void network as well: these clusters are embedded in
superclusters determined by rich clusters and populate filaments
between them.  We present a new catalog of superclusters of Abell
clusters out to a redshift of $z_{lim}=0.13$, a catalog of X-ray
clusters located in superclusters determined by Abell clusters, and a
list of additional superclusters of X-ray clusters.

\end{abstract}

  

\section{Introduction}

The formation of a filamentary web of galaxies and systems of galaxies
is predicted in any physically motivated model of structure formation
in the Universe (Bond, Kofman and Pogosyan 1996, Katz \etal 1996).
The largest relatively isolated density enhancements in the Universe
are superclusters of galaxies.  Observationally the presence of
superclusters and voids between them has been known since long ago (de
Vaucouleurs 1953, Abell 1958, Einasto, J\~oeveer, \& Saar 1980,
Zeldovich, Einasto \& Shandarin 1982, Oort 1983, Bahcall 1988).
Superclusters of galaxies and large voids between them form a
supercluster-void network of scale $100 - 120$~\Mpc\ ($h$ is the
Hubble constant in units of 100~km~s$^{-1}$~Mpc$^{-1}$).  The
supercluster-void network evolves from density perturbations of
similar wavelength (Frisch \etal 1995).  Superclusters correspond to
the density maxima, and the largest voids to the density minima of
perturbations of this scale, in a density field smoothed with a
Gaussian window of dispersion $\sim 8$~\Mpc (Frisch \etal 1995).  The
fact that superclusters are the largest physically well-defined
systems in the Universe is equivalent to the fact that they correspond
to the density perturbations of the largest relative amplitude.  On
these large scales the evolution of density perturbations is slow;
thus superclusters and their fine details grow from density
perturbations formed in the very early Universe.  In this way the
geometry of the supercluster-void network, as well as its fine
structure gives us information on the physical processes in the early
Universe.

The fine structure of superclusters with their galaxy and cluster
chains and filaments, and voids in-between, is presently quite well
studied.  The structure of the supercluster-void network itself is
known with much less accuracy.  Recently Einasto \etal (1994, 1997a,
1997c and 1997d, hereafter EETDA, E97a, E97c and E97d, respectively)
demonstrated the presence of a preferred scale of $120$\Mpc\ in the
distribution of rich clusters and superclusters of galaxies.  Although
several studies have found a maximum in the power spectra of galaxies
and clusters of galaxies at the same scale (Einasto \etal 1999a and
references therein), the shape of the power spectrum of clusters on
very large scales is not clear yet (Vogeley 1998, Miller and Batuski
2000).  The reason for this is simple: on scales larger than $\sim
100$~\Mpc\ the observational data are less complete.  On the other
hand, differences between cosmological models become significant only
on these larger scales, thus a better understanding of the real
situation is of great importance.

An independent line of evidence for the structure of the Universe on
large scales comes from the analysis of the CMB angular spectrum (de
Bernardis \etal 2000 and Hanany \etal 2000).  Fine structure of
temperature fluctuations on a degree scale has been detected; this
scale corresponds to a linear scale about 100~\Mpc; thus large scale
distribution of matter can be studied using combined CMB and optical
data.  These studies have caused increasing interest in the studies of
the clustering properties of matter on large scales.

So far superclusters have been determined using rich clusters of
galaxies from the catalogs by Abell (1958) and Abell, Corwin \& Olowin
(1989, hereafter ACO).  Abell samples of clusters of galaxies have
been used mainly for the reason that they form presently the largest
and deepest surveys of galaxy clusters available, containing more than
4000 clusters.  However, Abell clusters were found by visual
inspection of Palomar Observatory Sky Survey plates and the sample may
be influenced by various selection effects.  Selection effects change
the number of galaxies observed in clusters, and we can consider
observed catalogs of clusters as random selections from the underlying
true cluster sample using certain probabilities which represent
various selection effects.  The influence of these selection effects
can be studied by comparison of samples of clusters of galaxies
selected independently.  One of these optically selected independent
cluster samples is the catalog of clusters derived from scans with the
Automated Plate Measuring (APM) Facility (Dalton \etal 1997).  The
other possibility is to use samples of clusters selected by their hot
intracluster gas.  Hot gas accumulates in high-density regions; this
gas emits X-rays and can be detected by X-ray sensitive detectors
installed on satellites.  Resulting samples of X-ray selected clusters
of galaxies form independent samples selected from the same underlying
true cluster sample using different selection criteria.  In recent
years several catalogs of X-ray clusters have been published based on
ROSAT X-ray observations comprising data on several hundreds of these
objects.  These new catalogs have been used to investigate the
clustering properties of X-ray clusters recently. Usually these
studies analyze the correlation function on scales up to about
100~\Mpc\ (Romer \etal 1994, Abadi \etal 1999, Lee and Park 1999,
Moscardini \etal 1999a, Collins \etal 2000).  The clustering of the
X-ray clusters up to the same scales has been predicted theoretically
by Moscardini \etal (1999, 2000).

Another approach is to compile catalogs of superclusters of galaxies
and to study the distribution of clusters in superclusters.
Supercluster catalogues have been used for many purposes -- to
investigate the distribution of high-density regions in the Universe,
the large-scale motions in the Universe, the analysis of the
Sunyaev-Zeldovich effect (the scattering of the cosmic microwave
background radiation by hot gas in clusters and superclusters of
galaxies) in cosmic microwave background maps.  Examples of the last
type of analyses are Birkinshaw (1998), Refregier, Spergel \& Herbig
(2000), Kashlinsky \& Atrio-Barandela (2000). Diaferio, Sunyaev \&
Nusser (2000) propose that the presence of close large CMB decrements
may help to identify superclusters at cosmological distances.

The main goal of this series of papers is to compare the distribution
of Abell, X-ray selected and APM clusters of galaxies and to check how
well these cluster samples trace the properties of the underlying
true cluster distribution and the supercluster-void network. We
present an updated version of the supercluster catalog based on Abell
clusters, supercluster catalogs of X-ray and APM clusters, and a list
of X-ray clusters in superclusters determined by Abell clusters. We
compare the distribution of Abell, X-ray and APM clusters in different
environments.  The aim of this analysis is twofold: it gives us
information about the clustering properties of Abell, X-ray and APM
clusters; and independent evidence about how well different cluster
samples trace the distribution of high-density regions of the
Universe.  In the first paper of the series (this Paper) we compare
clustering properties of Abell and X-ray selected clusters in
superclusters.  In paper II we shall analyze the correlation function
of X-ray clusters and provide evidence for a characteristic scale of
$120$\Mpc\ in the distribution of X-ray clusters (Tago \etal 2001,
Paper II).  A similar comparison of Abell clusters and clusters found
from the Automatic Plate Measuring Machine (APM) catalog of galaxies
will be made by Einasto \etal (2001, Paper III).

The paper is organized as follows. In Section~2 we shall describe
cluster samples used and present an updated version of the catalog of
superclusters of Abell clusters.  In Section 3 we compile a list of
X-ray clusters in superclusters, analyze the distribution of Abell and
non-Abell clusters, calculate the fraction of X-ray clusters in
superclusters of different richness, and look for a relation between
X-ray luminosities of clusters with the richness of their parent
superclusters. In Section 4 we draw our conclusions.  In the Appendix
we present an updated version of the supercluster catalog based on
Abell clusters, and a list of X-ray clusters in superclusters and in
additional systems not present in the supercluster catalog.  The
catalog and both lists are also available electronically at web pages
of Tartu Observatory (www.aai.ee). There we also demonstrate 3-D
computer models and animations of the distribution of superclusters
and X-ray clusters.

\begin{figure*}[ht]
\vspace*{8.0cm}
\figcaption{Left panel: The multiplicity functions for Abell
clusters. The solid line shows the fraction of isolated
clusters as function of the neighborhood radius $R$; 
the short-dashed line shows the fraction of clusters in
medium-rich systems with a number of members from 2 to 31.  The dashed line
shows the fraction of clusters in very rich systems with at least 32
member clusters. Right panel: Supercluster multiplicities for a neighborhood
radius $R = 24$ \Mpc.  Isolated clusters are included for comparison.}
\includegraphics{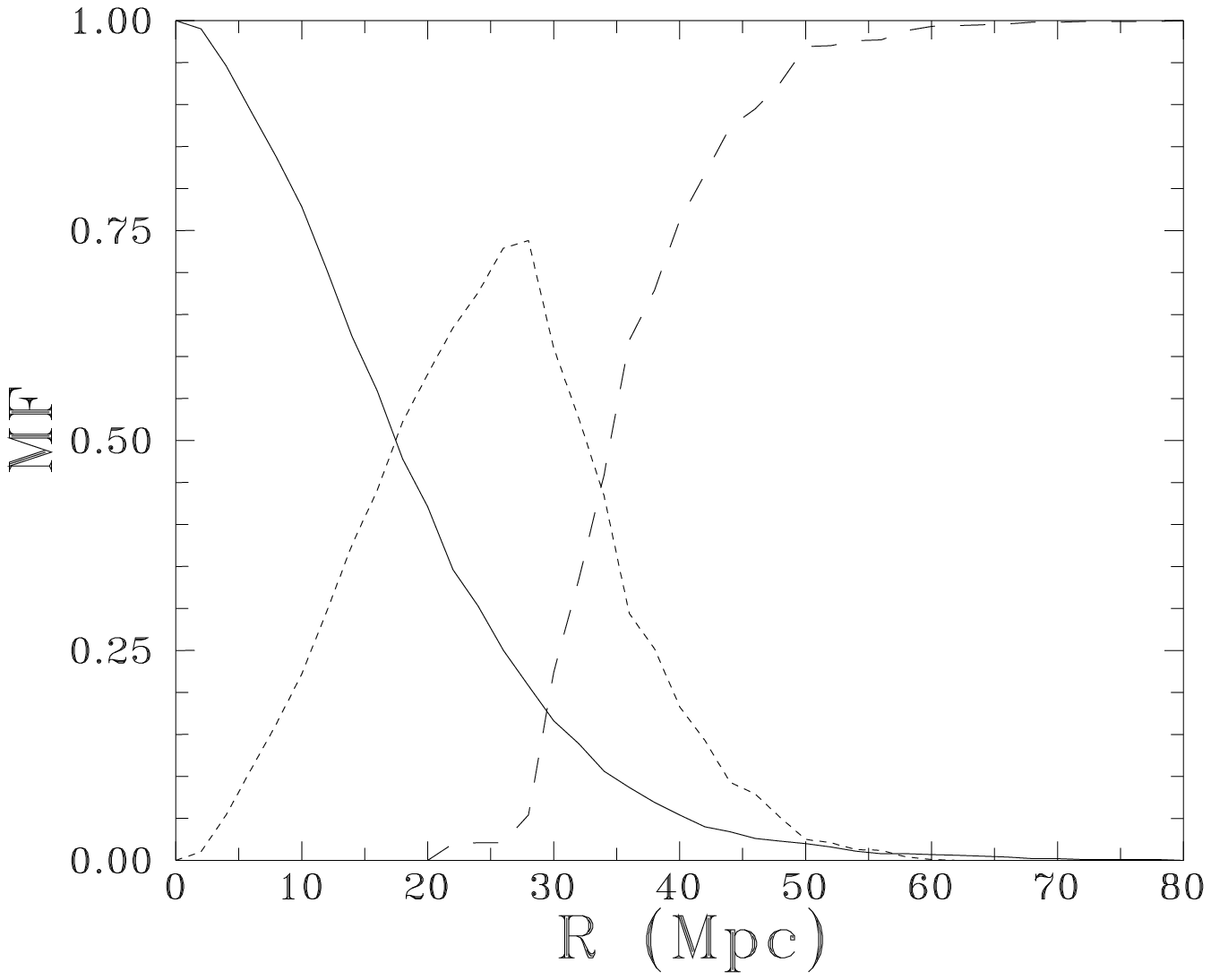}
\includegraphics{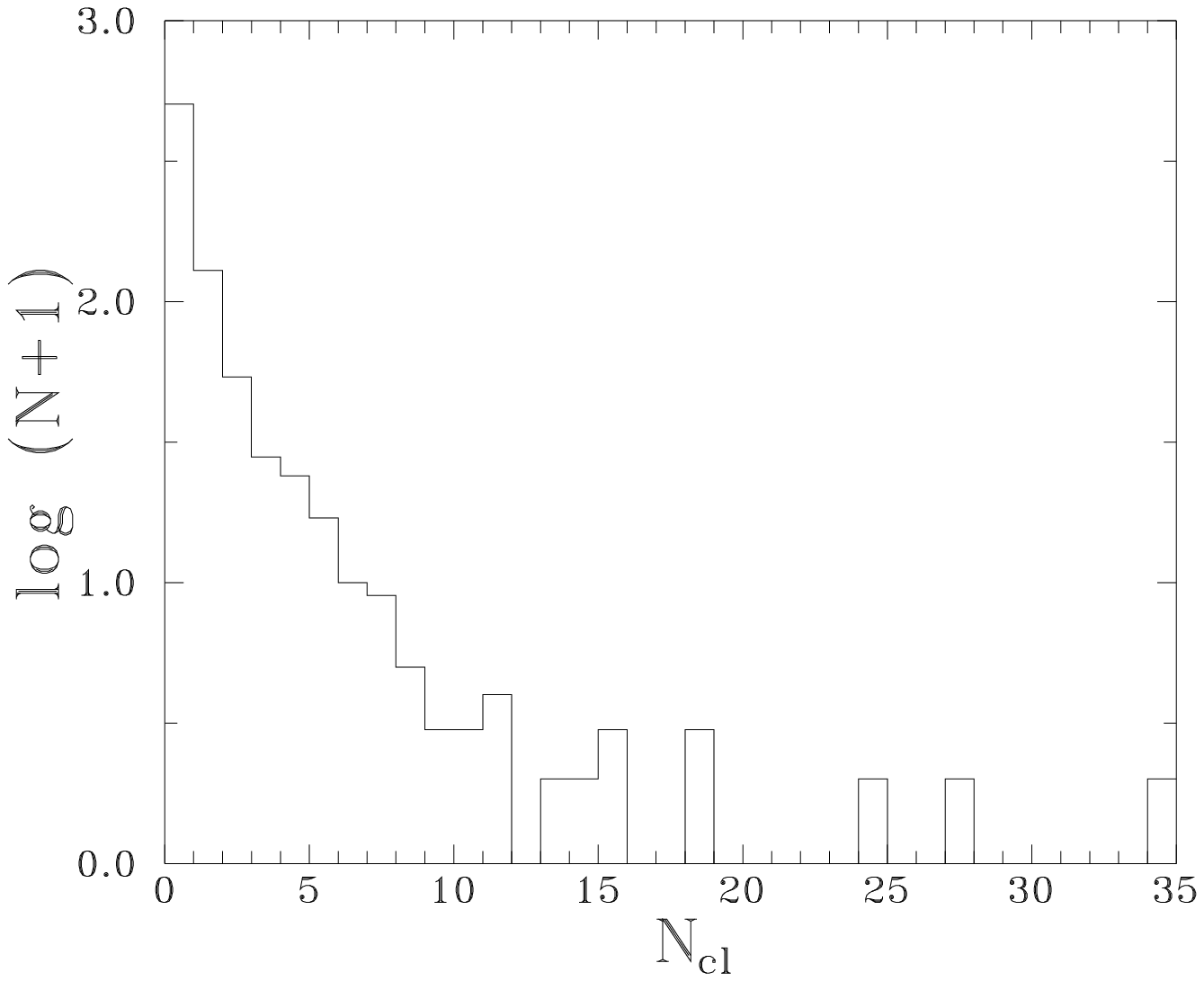}
\label{figure1}
\end{figure*}

\section{Data}

\subsection{Abell clusters}

For the present study we shall use the latest version (March 1999) of
the compilation of measured redshifts of Abell clusters described by
Andernach \& Tago (1998).  This compilation contains all known Abell
clusters with measured redshifts, based on redshifts of individual
cluster galaxies, and redshift estimates of the cluster according to
the formula derived by Peacock \& West (1992), for both Abell catalogs
(Abell 1958 and ACO).  We omitted from the compilation all
supplementary, or S-clusters, but included clusters of richness
class~0 from the main catalog.  From this general sample we selected
all clusters with measured redshifts not exceeding $z_{lim}=0.13$;
beyond this limit the fraction of clusters with measured redshifts
becomes small (selection effects in the Abell cluster sample up to
redshift $z_{lim}=0.15$ shall be studied in Paper III).  If no
measured redshift was available we applied the same criterion for
estimated redshifts. Our sample contains 1662 clusters, 1071 of which
have measured redshifts.  We consider that a cluster has a measured
redshift if at least one of its member galaxy has a measured redshift.
In cases where the cluster has less than three galaxies with measured
redshifts, and the measured and estimated redshifts differ more than a
factor of two ($|\log( z_{meas}/z_{est})| > 0.3$), the estimated
redshift was used.  In the case of superimposed clusters or component
clusters (A,B,C etc) with comparable number of measured redshifts, we
used only the cluster which better matches the estimated redshift.

Distances to clusters have been calculated using the following formula
(Mattig 1958):
$$
r = {c \over {H_0 q_0^2}} {{q_0 z + (q_0-1)(\sqrt{1+2q_0z} -1)}
\over {1+z}};
\eqno(1)
$$
where $c$ is the velocity of light; $H_0$ -- the Hubble parameter; and
$q_0$ -- the deceleration parameter.  We use $H_0 =
100~h^{-1}$~km~s$^{-1}$~Mpc$^{-1}$, and $q_0=0.5$.

\subsection{Superclusters of Abell clusters}

On the basis of the Abell cluster sample we constructed a list of
superclusters of Abell clusters using a friends-of-friends (FoF)
algorithm described in detail by EETDA and E97c.  Clusters are
assigned to superclusters using a certain neighborhood radius so that
all clusters in the system have at least one neighbor at a distance
not exceeding this radius.  The neighborhood radius to assign clusters
to superclusters should be chosen in accordance with the spatial
density of the cluster sample.  Also, we define the multiplicity of a
supercluster (supercluster richness), N$_{CL}$, as the number of its
member clusters. Superclusters are divided into richness classes as in
E97c: poor superclusters (number of members $N_{CL}= 2,~3$), rich
superclusters ($4 \leq N_{CL} \leq 7$), and very rich superclusters
($N_{CL} \geq 8$).

In Figure~1 (left panel) we show the fraction of clusters in systems
of different multiplicity for a wide range of neighborhood radii for
the Abell cluster sample.  At small radii all clusters are
isolated. With increasing neighborhood radius some clusters form
superclusters of intermediate richness. In Figure~1 we plot the
fraction of clusters in superclusters of richness $2 \le N_{CL} \le
31$. At larger radii extremely large superclusters with multiplicity
$N_{CL} \ge 32$ start to form.  By further increasing the neighborhood
radius superclusters begin to merge into huge conglomerates; finally
all clusters percolate and form a single system penetrating the whole
space under study.  In order to obtain superclusters as the largest
still relatively isolated systems we must choose a neighborhood radius
smaller than the percolation radius.  The appropriate neighborhood
radius is the radius which corresponds to the maximum of the fraction
of clusters in systems of intermediate richness.  Beyond this radius
very large systems start to form, as seen from Figure~1 (see also
EETDA and E97c). For Abell clusters the appropriate neighborhood
radius to select systems is $24$~\Mpc.  We shall apply the same radius
to the samples of X-ray clusters in order to determine which non-Abell
X-ray clusters are the members of superclusters of Abell clusters, as
well as to detect additional superclusters of non-Abell X-ray
clusters.

For the present study we update the supercluster catalog and determine
systems up to redshifts $z = 0.13$. This larger redshift limit was
used in order to include several distant rich superclusters whose
members have measured redshifts and which also contain X-ray clusters,
e.g.  the Draco-Ursa Majoris supercluster with 14 member clusters.
The new Abell supercluster catalog contains 285 superclusters with at
least 2 member clusters, 31 of them are very rich superclusters with
at least 8 members. The catalog of superclusters of Abell clusters is
given in the Appendix (Table A1). In Figure~1 (right panel) we plot
supercluster multiplicities for this catalog.  In the present study
this supercluster catalog was used as a reference to look for X-ray
clusters in superclusters.

\subsection{X-ray selected cluster samples}

The ROSAT observations were made with the Position Sensitive
Proportional Counter during the ROSAT All-sky Survey (RASS) in 1990
and 1991 (Tr\"umper 1993). After that the so-called Guest Observers
(GO) four-year observing program was completed.

On the basis of RASS data several catalogs of X-ray selected clusters
of galaxies were prepared.  In the present paper we shall use the
following samples of X-ray clusters:

i) clusters from the all-sky ROSAT Bright Survey of high Galactic
latitude RASS sources. A detailed description of the data is given in
Voges \etal 1999, and the catalog of X-ray clusters, AGNs, galaxies,
small groups of galaxies and other objects in Schwope \etal 2000.  We
shall refer to this sample as RBS.

ii) ROSAT PSPC observations of the richest ($R \geq 2$) ACO clusters
(David, Forman and Jones 1999, hereafter DFJ);

iii) a flux-limited sample of bright clusters from the Southern sky 
(de Grandi \etal 1999, see also Guzzo \etal 1999);

iv) the ROSAT brightest cluster sample (Ebeling \etal 1998, BCS) from
the Northern sky.

Redshifts are available for all the clusters.  

The ROSAT Bright Survey is the only available all-sky survey of X-ray
clusters. Objects in this survey have been selected at Galactic
latitudes, $|b| > 30^{\circ}$, with PSPC count rate larger than 0.2~
s$^{-1}$ and flux limit $2.4\times10^{-12}$~erg~cm$^{-2}$~s$^{-1}$ in
the hard energy band ($0.5 - 2.0$~keV).  For our analysis we selected
clusters with measured redshifts up to $ z = 0.13$ -- the redshift
limit of the catalog of superclusters of Abell clusters (see above).
Altogether, this sample comprises 203 clusters, including 40 non-Abell
clusters.  We shall refer to this cluster sample as the ``RBSC''
sample; for cluster numbers we use RBS numbers as given in Schwope et
al. (2000).

Further, we use the list of the richest ($R \geq 2$) Abell clusters
detected with ROSAT PSPC observations (DFJ).  This catalog contains
data on the clusters of galaxies observed during the GO phase of the
ROSAT mission.  The main advantage of these observations is longer
exposure time (typically 10 000 seconds) than in the RASS (400
seconds). However, the sky coverage of this compilation is far less
than that of RBSC catalog since the latter clusters were found in
targeted and serendipitous observations.  For the method to calculate
X-ray fluxes we refer to DFJ.  Up to distances $z = 0.13$ this sample
contains 52 clusters.  We shall denote this sample as DFJ.

The Brightest Cluster Sample (BCS, Ebeling \etal 1998) covers the
Northern sky ($\delta > 0^{\circ}$) at Galactic latitudes $|b| >
20^{\circ}$ in the broad energy band ($0.1 - 2.4$~keV).  The lower
flux limit for sample was 4.4 $10^{-12}$ergs~ cm$^{-2} s^{-1}$.
Ebeling \etal developed the VTP (Voronoi Tessellation and Percolation)
algorithm to determine X-ray fluxes of extended sources of arbitrary
shapes.  Up to $z = 0.13$ this sample contains 141 clusters, including
46 non-Abell clusters.  We shall denote this sample as BCS.

The flux-limited sample of bright clusters of galaxies from the
Southern sky by de Grandi \etal (1999) is selected at galactic
latitudes $|b| > 20^{\circ}$, the declination $\delta < 2.5^{\circ}$,
and the flux limit in the hard band ($0.5 - 2.0$~keV) was $3 - 4
\times 10^{-12}$~ erg~cm$^{-2}$~s$^{-1}$.  In their study the
so-called Steepness Ratio Technique was used to determine X-ray
fluxes.  Up to $z = 0.13$ this sample contains 101 clusters, 34 of
which are non-Abell clusters.

We shall discuss the completeness and selection effects of Abell and
X-ray clusters in Paper II. In general, at distances larger than
approximately 250~\Mpc\ the samples of X-ray clusters are rather
diluted due to the fixed flux limit; on larger distances X-ray clusters
have been used in the present paper for lists of supercluster members
only (and not for correlation analysis in Paper II).

\begin{figure*}[ht]
\vspace*{16.0cm} \figcaption{The distribution of X-ray clusters
(filled symbols, supercluster members) and Abell clusters (open
circles) in supergalactic coordinates. In order to avoid overcrowding
of the figure we plot only clusters from very rich superclusters) in
supergalactic coordinates. In each panel we plot Abell clusters and
X-ray clusters from one sample. X-ray samples are plotted as follows.
Upper left panel: $RBS$ sample.  Here we plot also members of
additional systems (squares, Table B2), and isolated non-Abell
clusters (triangles); upper right panel: $DFJ$ sample; lower left
panel: $BCS$ sample, and lower right panel: sample by de Grandi \etal
(1999). The extent of all panels in supergalactic X coordinate is
600~\Mpc\ } 
\includegraphics{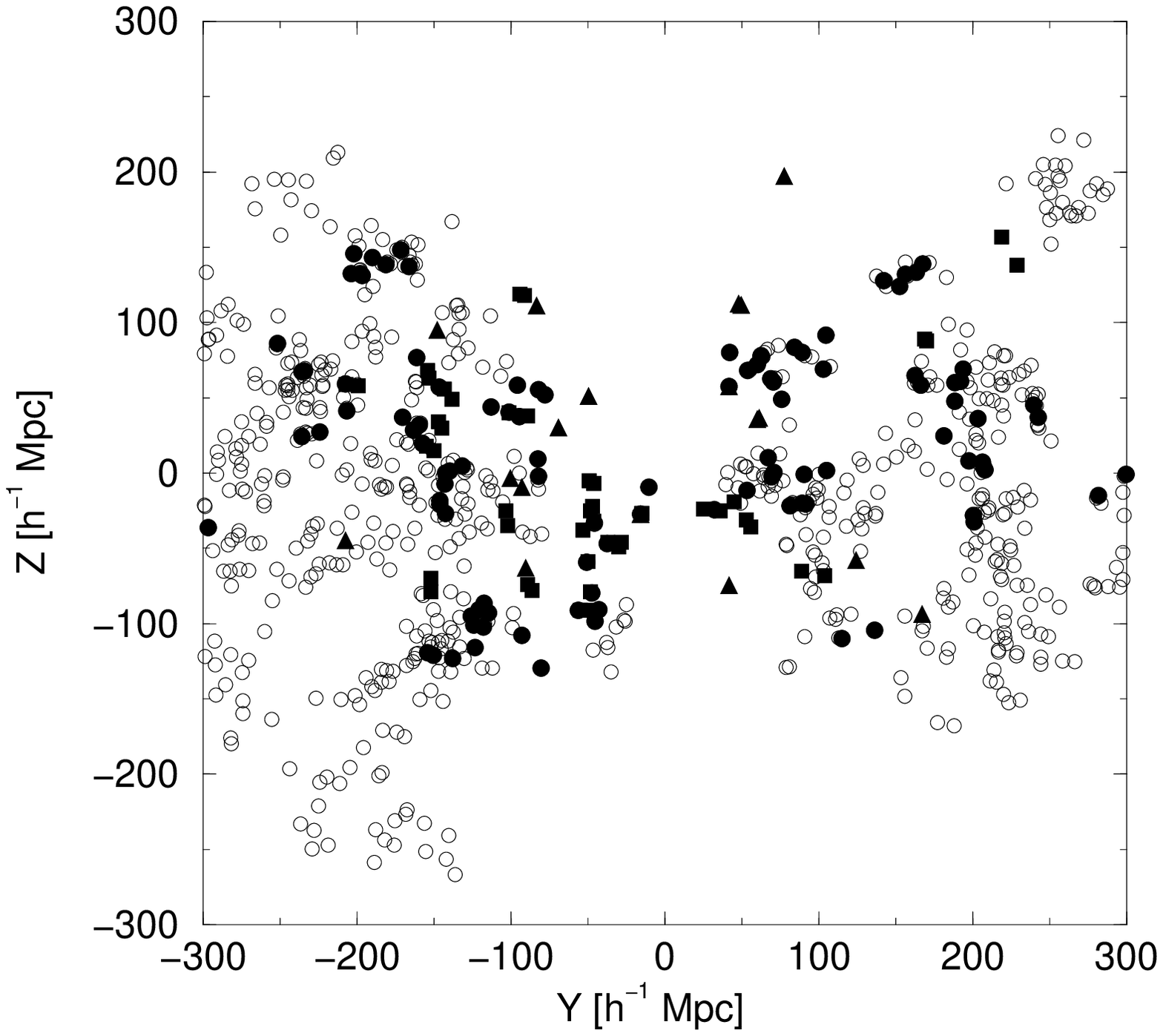} 
\includegraphics{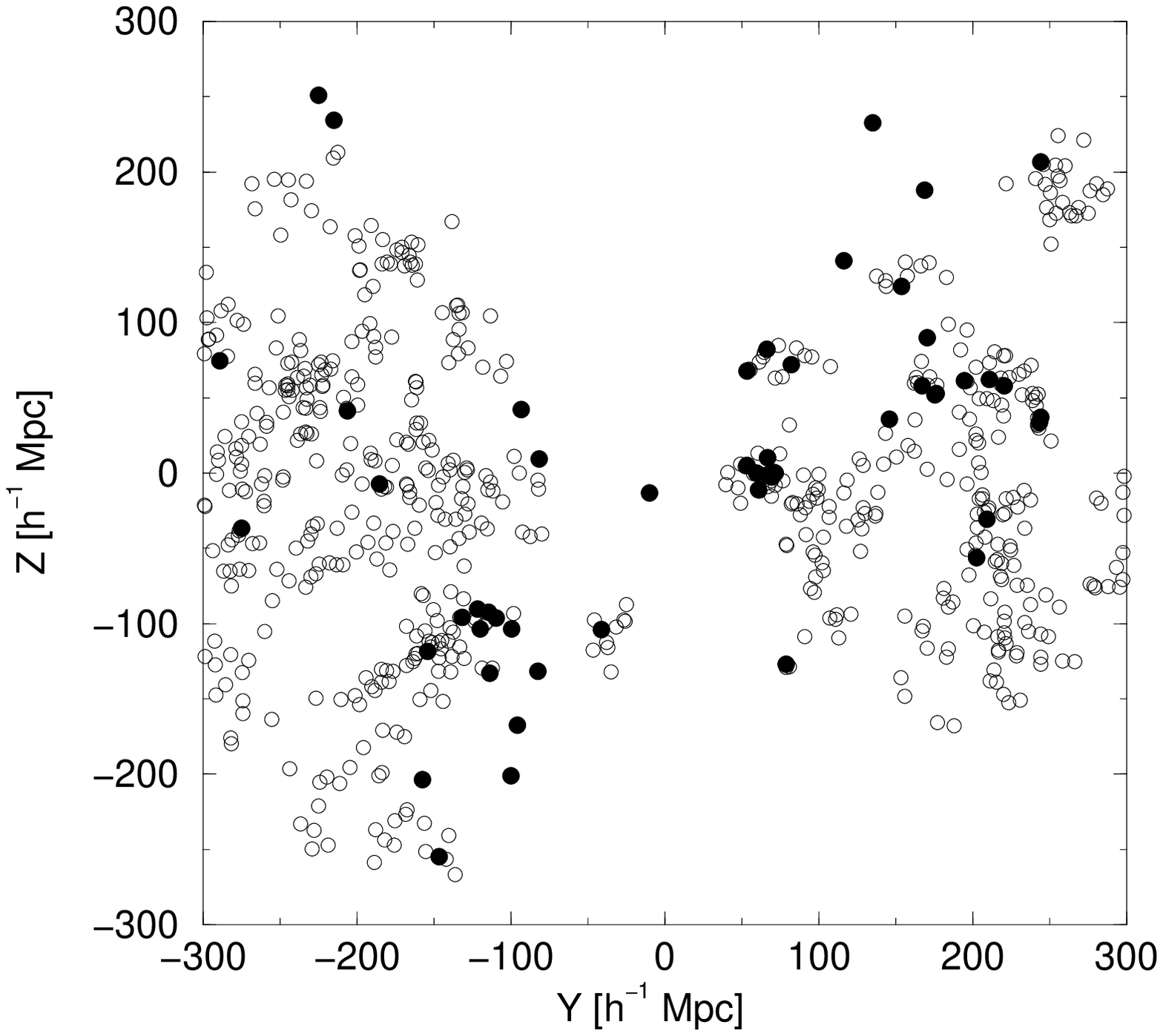} 
\includegraphics{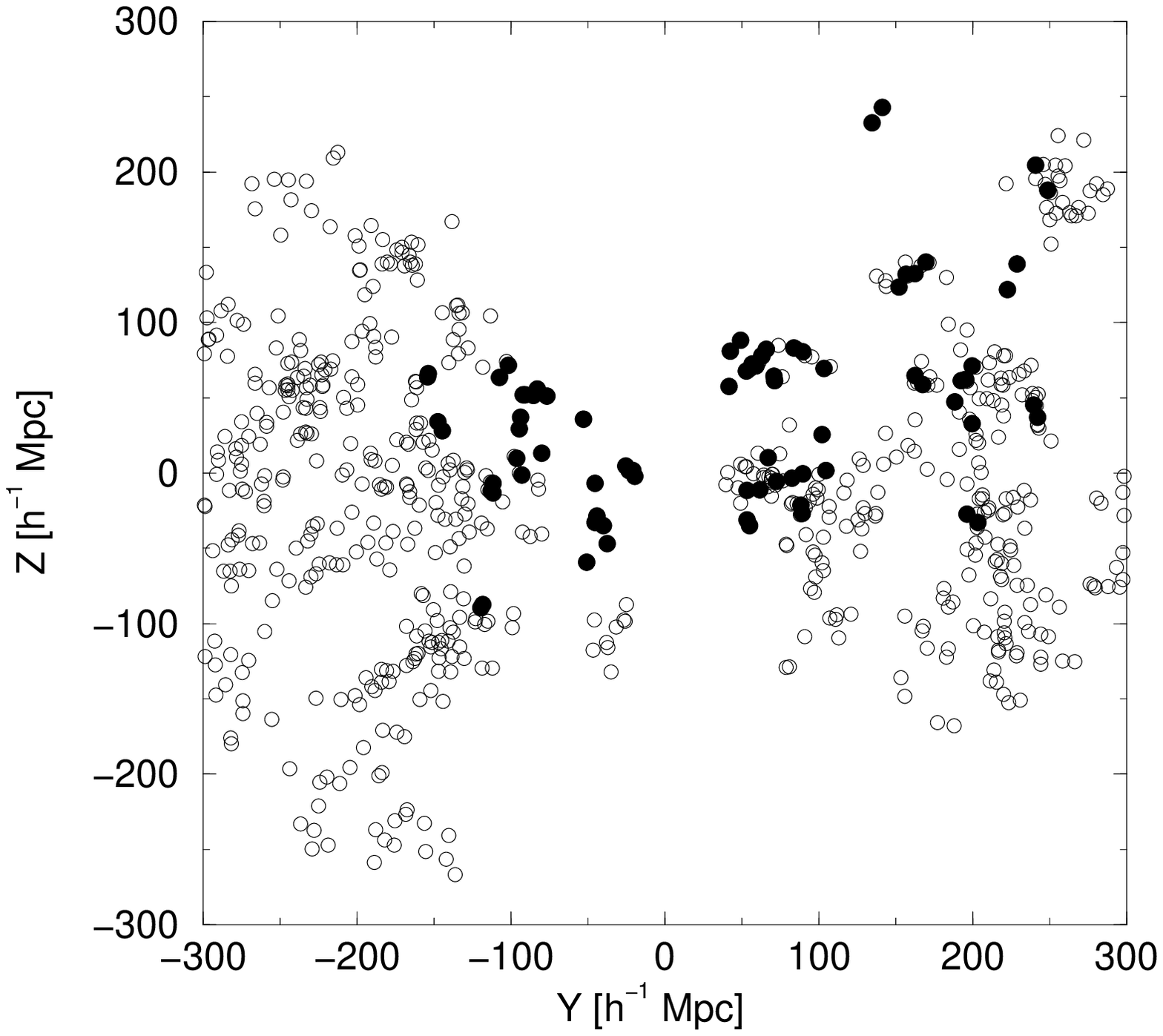} 
\includegraphics{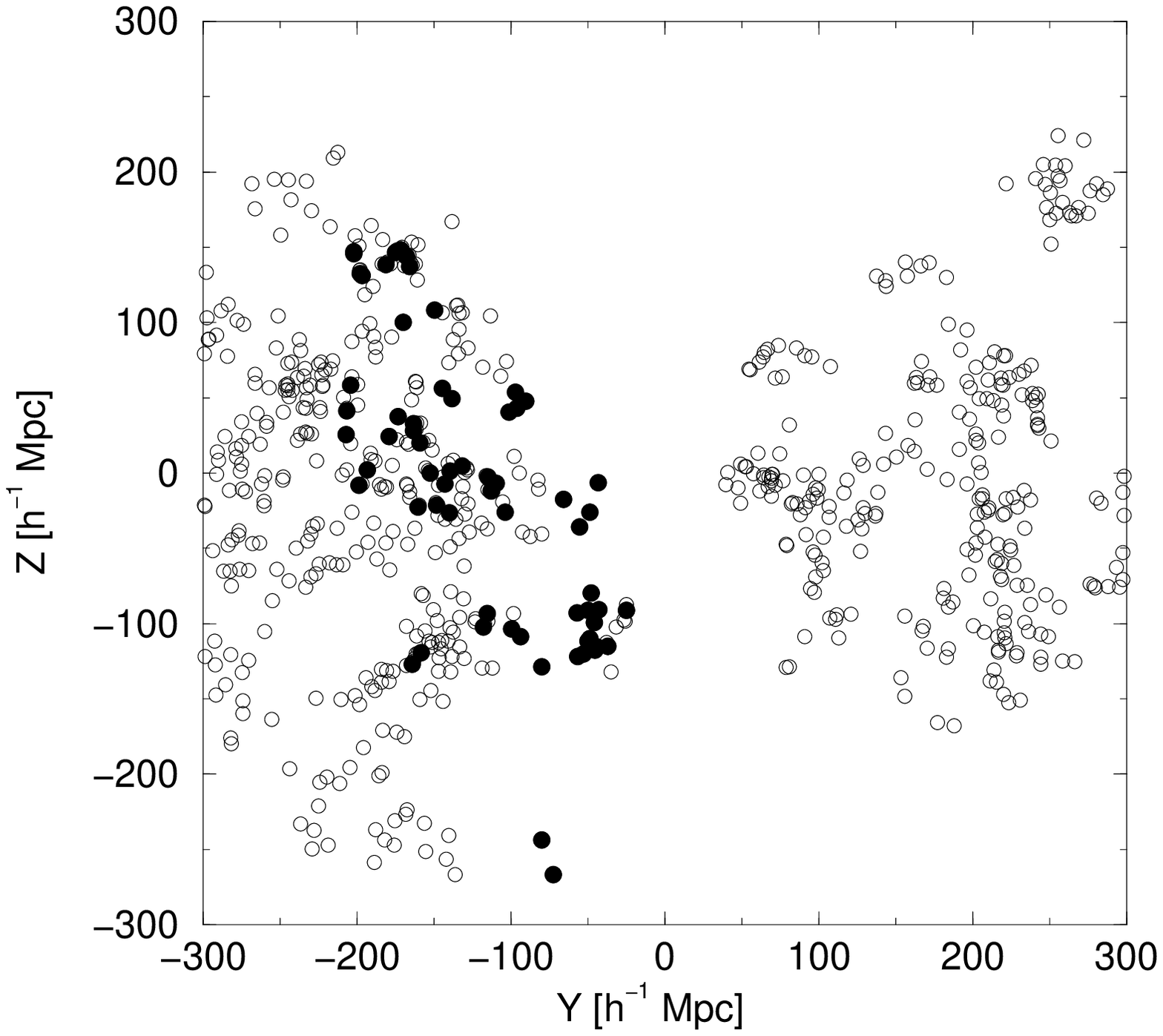}
\label{figure2}
\end{figure*}

\begin{table*}
\begin{center}
\caption[dummy]{Fraction of X-ray clusters in superclusters of different
  richness }
\label{tab:param}
\begin{tabular}{lrccccc}
\\
\hline
\\
Supercluster richness  & $N_{A}$  &  $F$ &\multispan4  $N_{X-ray}$ \\
  &   &  & $N_{A}$ &  $F_{A}$ &$N_{nA}$ &  $F_{nA}$\\
\hline
\\
scl members &  1256  &  & 182 & &   68 &   \\ 
poor ($2\leq N _{cl}\leq 3$) &  513  &   41$\%$  & 47   & 26$\%$  
 &  18 &  26$\%$\\
rich ($4\leq N _{cl}\leq 7$)  &  370  &   29$\%$  &   59 & 32$\%$  
 &  20 &  30$\%$\\
very rich ($N _{cl}> 8$)  &  373  &   30$\%$  &   76 & 42$\%$   & 
 30 &  44$\%$\\
\\
\hline
\end{tabular}
\end{center}
\end{table*}

\section{X-ray clusters in superclusters}

In this Section we compile a list of X-ray clusters that belong to the
superclusters derived from Abell clusters as listed in Table~A1. In
addition, we searched for systems consisting of non-Abell X-ray
clusters and determine their location with respect to the
supercluster-void network.  We also calculate the fraction of X-ray
clusters in superclusters of various richness and investigate the
possible correlation between cluster X-ray luminosities and
supercluster richnesses.

\subsection{A list of X-ray clusters in superclusters}
In Table B1 we present a list of X-ray clusters in superclusters of
Abell clusters presented in Table A1.  Abell clusters from X-ray
catalogs were included by comparison of the catalogs of X-ray clusters
with the supercluster catalog.  In order to include non-Abell X-ray
clusters we searched for superclusters that contain X-ray clusters in
two ways.  First, we added non-Abell X-ray clusters to our Abell
cluster catalog and applied the FoF algorithm to this combined
catalog.  Second, we applied the FoF algorithm to each catalog of
X-ray clusters separately. In both cases we used the same neighborhood
radius, $R = 24$~\Mpc\, as in the case of Abell clusters. The second
procedure was used to check whether X-ray clusters that are
supercluster members form systems by themselves also.  Additionally,
for some superclusters this second procedure detects outlying Abell
clusters as members of superclusters that are not listed in Table A1
(mainly due to small differences in redshift measurements).  In the
case of X-ray clusters identified as Abell clusters this double
procedure gives us additional evidence about the reliability of the
superclusters found by optical surveys.

Non-Abell clusters that were found to be members of superclusters of
Abell clusters (Table~A1) were considered as members of these systems.
However, their membership has to be checked carefully. The
superclusters of Abell clusters were defined as the largest still
relatively isolated systems.  In some cases non-Abell clusters (poor
clusters of galaxies) really belong to the superclusters, but in other
cases non-Abell clusters actually form a bridge of poor clusters that
connect superclusters of Abell clusters. Therefore, the actual
location of each non-Abell cluster that was connected to some
supercluster according to the FoF algorithm was checked separately.
We shall mention below the cases when clusters formed filaments
connecting superclusters, rather than forming new members of a single
supercluster.

We note that in most cases when a supercluster contains more than one
X-ray cluster, these X-ray clusters themselves form a supercluster at
the neighborhood radius $R = 24$~\Mpc. Therefore Table B1 lists
superclusters of X-ray clusters as well. Only in a few cases of very
elongated superclusters it happened that some X-ray members of the
system remained as separate systems so that the supercluster was split
into smaller systems.  The supercluster number in the column 1 of Table B1
correspond to supercluster numbers from the catalog in Table A1.

The use of combined (X-ray and optical) data to determine X-ray
clusters in superclusters was very fruitful.  In our catalog of
superclusters containing X-ray clusters (Table~B1) there are 99
superclusters.  Of these superclusters 53 contain only one member as
an X-ray cluster.  These X-ray clusters would be isolated if we would
use data on X-ray clusters only; actually they are members of
superclusters.  Such an approach could be useful in the analysis of
systems of X-ray selected AGNs, as mentioned also in Tesch and Engels
(2000).

In Table B2 we list additional superclusters that contain non-Abell
clusters.  In most cases these systems are pairs of Abell and
non-Abell X-ray clusters.  Most Abell clusters in these superclusters
were isolated if only Abell clusters were used in supercluster
search. We shall denote these superclusters as $SCLX$ + supercluster
number from Table B2.

\subsection{Comments on individual superclusters}

{\it The Hercules supercluster (SCL 160)} at a distance of about
$100$ \Mpc\ contains the largest number of X-ray clusters -- 14,
including 7 non-Abell clusters. All of them are probably true
supercluster members.

{\it The Shapley supercluster (SCL 124)} at a distance of about
$130$ \Mpc\ contains 9 X-ray clusters, only one of them is a non-Abell
cluster. In this supercluster X-ray emission has been detected also
from filaments of galaxies connecting individual clusters (Bardelli
\etal 1998 and references therein, Kull and B\"ohringer 1999, Ettori
\etal 1997).

{\it The Horologium-Reticulum supercluster (SCL 48)}, one of the
richest superclusters in the Southern sky, is also very rich in X-ray
clusters, containing 11 X-ray clusters; only one of them is a non-Abell
cluster.
We note that the number of optically very rich X-ray clusters from
the compilation by DFJ is the largest in the last two superclusters,
in the Shapley and in the Horologium-Reticulum superclusters, both
containing six  X-ray clusters.

{\it The supercluster SCL 170} is very interesting.  According to the
data used in our study this supercluster contains only one X-ray
cluster -- A2312.  Actually this supercluster is one of the richest in
X-ray clusters -- it is the North Ecliptic Pole (NEP) supercluster
(Mullis 1999, Mullis \etal 2000) that contains approximately 15 X-ray
clusters. In the NEP survey the X-ray flux limit was lower than in the
catalogs used in our study and thus contains fainter X-ray clusters
than those catalogs.  This example shows that our list of X-ray
clusters in superclusters compiled on the basis of the X-ray brightest
cluster catalogs is preliminary, containing the X-ray brightest
supercluster members only.

{\it The Pisces supercluster} contains 10 X-ray clusters, 4 of which
are non-Abell clusters. However, our analysis shows that actually
these poor clusters belong to a filament that connects the Pisces
supercluster and superclusters 211 and 215.

Poor X-ray clusters connect the {\it Coma} and the {\it Leo}
superclusters (SCL 117 and 93), the {\it Sculptor} supercluster (SCL
9) and SCL 220 (see also Paper II), SCL 126 and 136, and SCL 212 and
297.  These cases confirm that poor X-ray clusters trace the
supercluster-void network determined by Abell clusters. X-ray clusters
either belong to superclusters themselves or they form filaments between
them.

Additional superclusters of X-ray clusters from Table~B2, being
located in filaments between superclusters, also trace the
supercluster-void network. Several of these systems ($SCLX$ 7, 9, and
12) border the Southern and Northern Local Supervoids (EETDA).  $SCLX$
9 contains one of the X-ray brightest Abell clusters, A496, see above,
and in addition to poor clusters this system harbors two X-ray
detected AGNs, RBS 550 and RBS 556.  X-ray detected AGNs from the RBS
catalog connect $SCLX$ 4 and 7 from Table~B2. This joint system
contains 11 AGNs and 7 X-ray selected clusters, including 3 Abell
clusters and one QSO (QSO 0351+026).

\begin{figure*}[ht]
\vspace*{16.0cm} 
\figcaption{X-ray luminosities for clusters in
superclusters of different richness and for isolated clusters (in
units of $10^{43}$erg~s$^{-1}$); clusters of the highest X-ray
luminosities are indicated below in parenthesis. X-ray samples are
plotted as follows: upper left panel: RBS sample (A2142, A2029, A401);
upper right panel: DFJ sample (A2142, A2029, A478); lower left panel:
BCS sample (A2142, A2029, A478); lower right panel: sample by de
Grandi \etal. (A3266, A3186, A3827).}  
\includegraphics{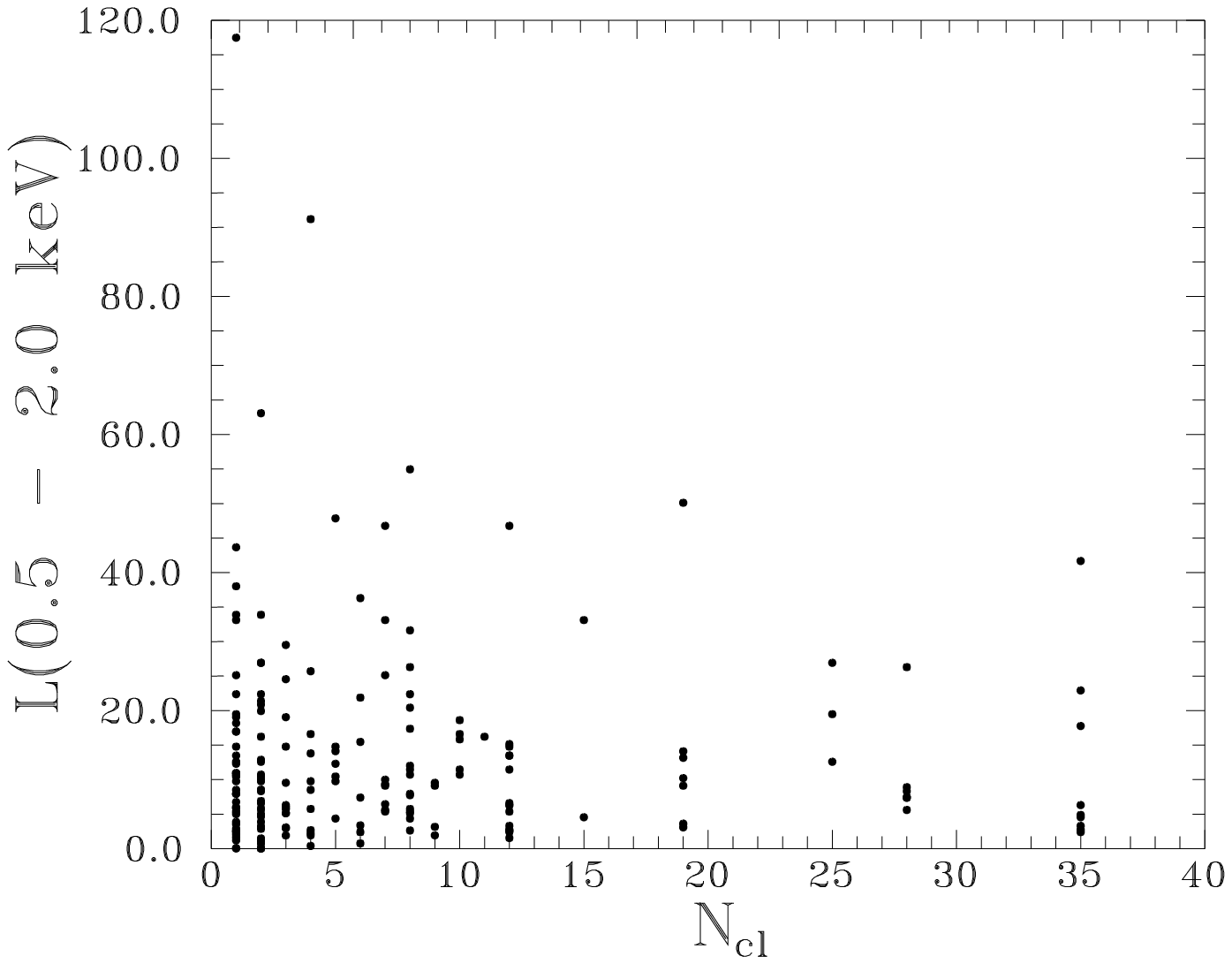} 
\includegraphics{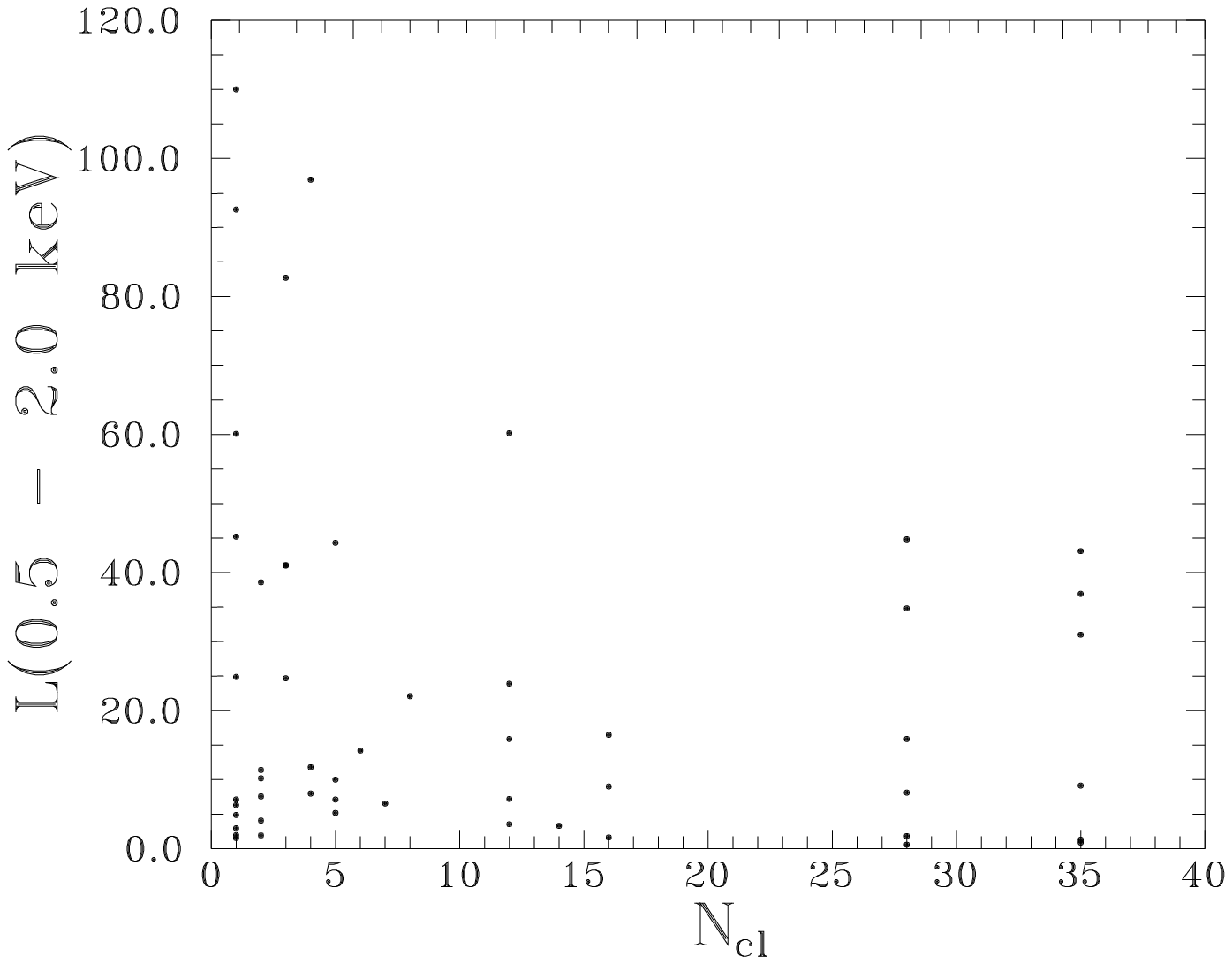} 
\includegraphics{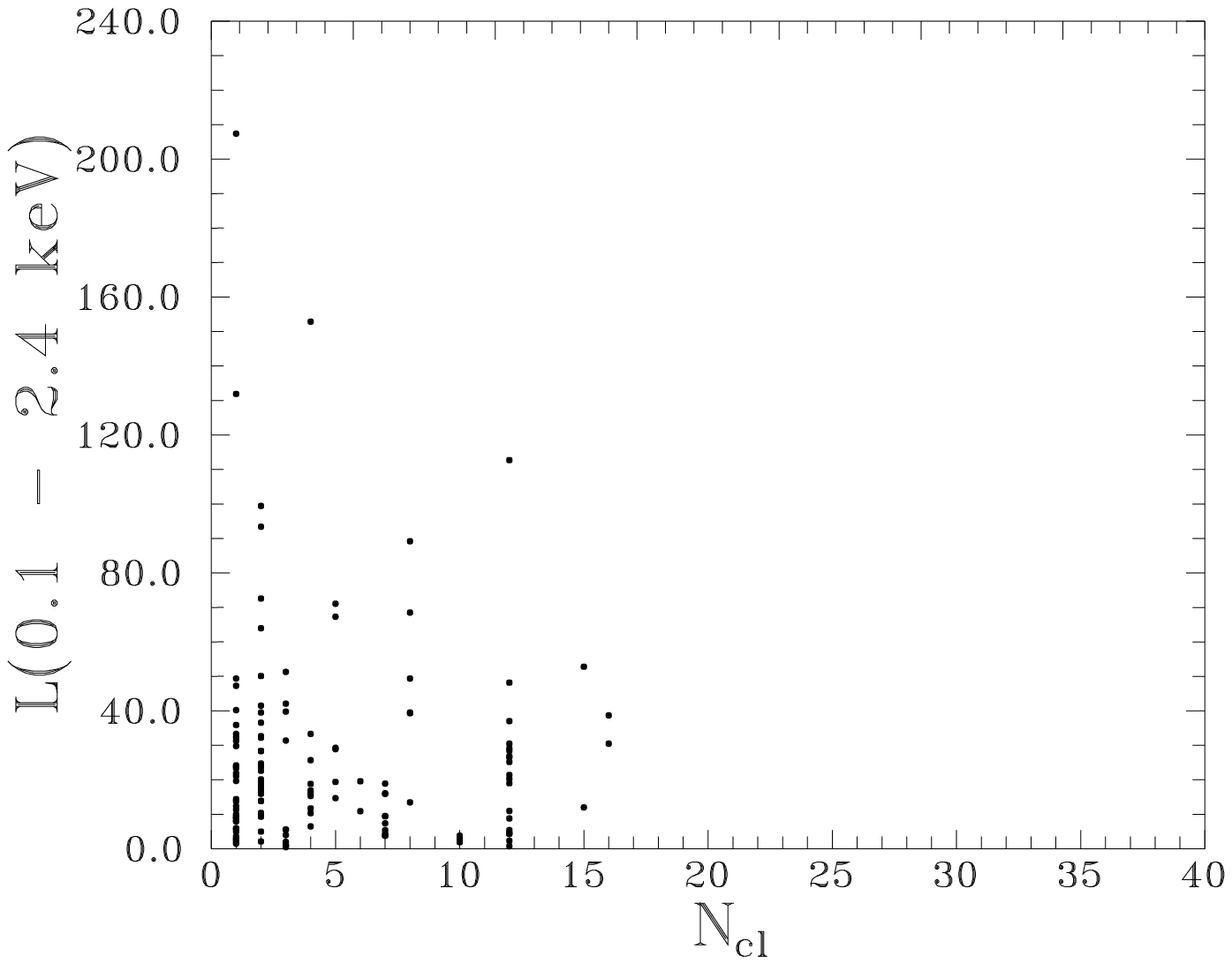} 
\includegraphics{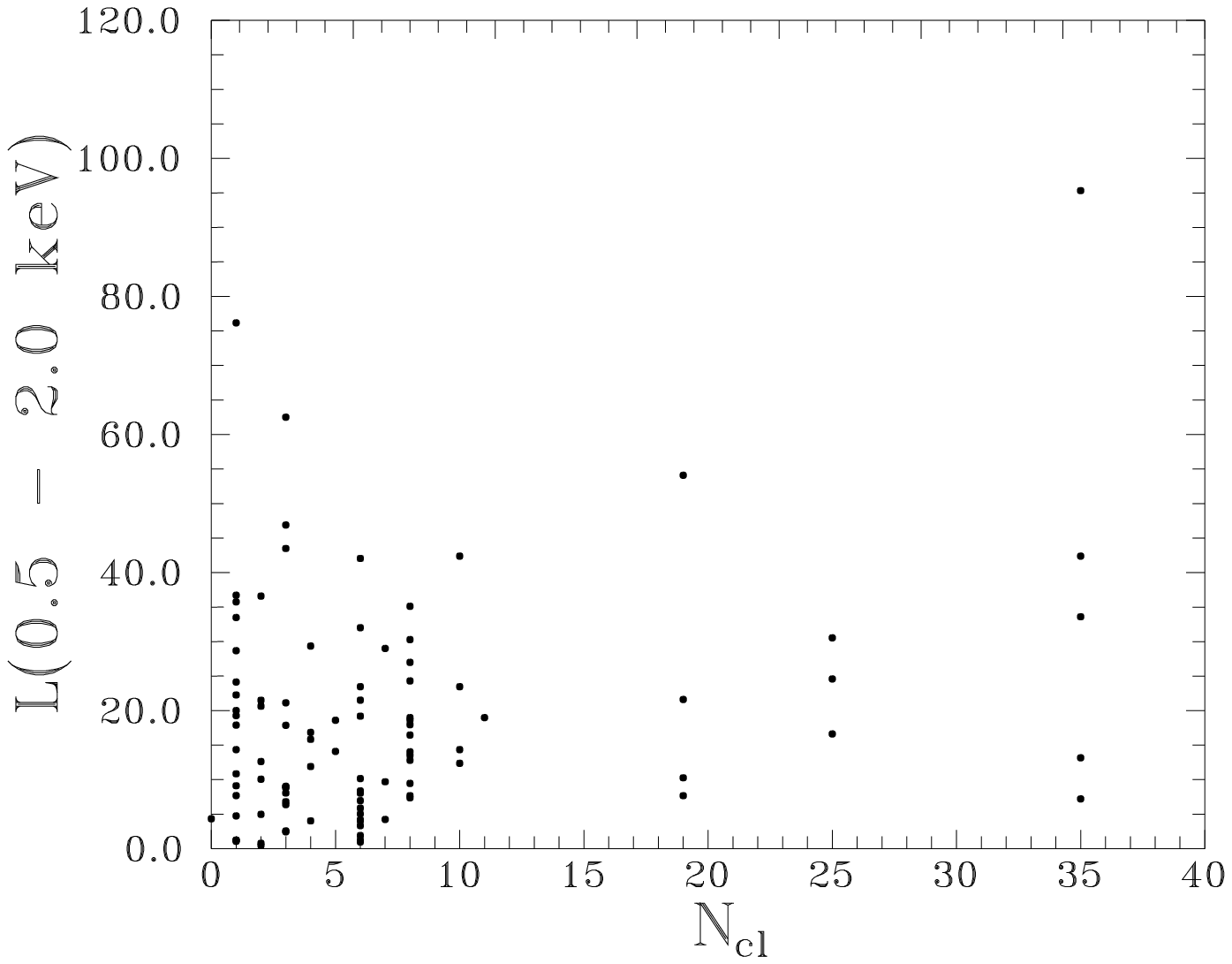}
\label{figure3}
\end{figure*}

In EETDA we showed that isolated Abell clusters are located close to
the superclusters and do not fill in the voids between superclusters.
Our present analysis shows additionally that most of the isolated poor
X-ray clusters that do not have neighbors at $R \geq 24$ \Mpc\ are
located in filaments between superclusters or on the borders of 
Southern and Northern Local voids.

In Figure~2 we plot the distribution of X-ray clusters and Abell
clusters that belong to very rich superclusters.  We see that the
structures delineated by optical and X-ray clusters coincide and we
can see a pattern of superclusters and voids.  The supercluster-void
network is more clearly seen in three-dimensional animations from our
web page, www.aai.ee.

In this Figure we plot also clusters from additional superclusters
(Table B2), as well as the location of isolated non-Abell clusters.
Many of them are located near the zone of avoidance where cluster
catalogs tend to be incomplete and superclusters cannot be determined.

\subsection{Fraction of X-ray clusters in superclusters}

After compiling the list of X-ray clusters in superclusters we
calculate the fractions of these clusters in superclusters of various
richness (Table~1).  Superclusters are divided into richness classes
as in E97c: poor superclusters (number of members $N_{CL}= 2, 3$),
rich superclusters ($4 \leq N_{CL} \leq 7$), and very rich
superclusters ($N_{CL} \geq 8$). Additionally, we give the fraction of
isolated X-ray clusters.

Table~1 shows that the fraction of X-ray clusters in superclusters
increases with increasing supercluster richness.  The
Kolmogorov-Smirnov test confirms that the zero hypothesis (the
distributions of optical and X-ray clusters in superclusters of
various richness are statistically identical) is rejected at the 99\%
confidence level.  In total, about one third of all superclusters and
23 of 29 very rich superclusters contain X-ray clusters.  About $25$\%
of Abell clusters are isolated at the neighborhood radius $R = 24$
\Mpc.  In contrast, only about $15\%$ of X-ray clusters are isolated
at this radius.

We note that various surveys used in the present study show a similar
tendency -- the increase of the fraction of X-ray clusters with
supercluster richness. However, the exact percentages of X-ray
clusters in systems of various richness are somewhat different due to
the differences between samples.  For example, due to the sky coverage
limits the fraction of isolated clusters is relatively high in the BCS
sample (25\% of poor clusters in this sample are isolated, see also
Paper II).  Also, due to the incompleteness of X-ray cluster catalogs
at large distances these fractions should actually be taken as lower
limits: at distances larger than $R = 275$~\Mpc\ there are only five
supercluster with more than one X-ray member cluster, and over 20
superclusters containing one X-ray cluster only.  However, test
calculations with smaller, statistically more complete subsample from
RBSC catalog in which clusters were selected up to the distance $R =
250$~\Mpc\ (Paper II) confirm that the fraction of X-ray clusters in
rich superclusters is higher than in poor superclusters.

\subsection{X-ray luminosities of clusters in superclusters 
of different richness}

In Figure~3 we plot X-ray luminosities for clusters in superclusters
of different richness in units of $10^{43}$~erg~s$^{-1}$.  X-ray
luminosities are calculated differently in the various X-ray cluster
catalogs.  In some catalogs the broad energy band (0.1 -2.4 keV) is
used (e.g. the BCS sample), while others are based on the hard energy
band (0.5 - 2.0 keV).  Also, different methods are used to determine
the total X-ray flux of extended sources. As a result, the X-ray
luminosities for various cluster samples are not directly comparable,
particularly in the case of clusters with complicated morphology.
However, our aim is to see whether cluster X-ray luminosities are
correlated with host supercluster richness, and for that purpose we
may simply plot X-ray luminosities for each sample separately.

Figure~3 shows that some clusters of very high X-ray luminosity are
located in superclusters of low multiplicity. Since Figure~3 does not
show any other clear correlation between cluster X-ray luminosities
and their host supercluster richness we think that it is preliminary
to draw quantitative conclusions from this finding.  Instead, we
describe shortly the locations and properties of the brightest X-ray
clusters.

The cluster with the highest X-ray luminosity in the Northern sky is
A2142. This cluster is isolated and located in the low-density
filament of clusters connecting the Corona Borealis and the Bootes A
superclusters (SCL 158 and 150).  Evidence was found for an ongoing
merging of two subclusters in this cooling flow cluster (Markevitch
\etal 2000 and references therein, and White, Jones and Forman 1997).

The second brightest X-ray cluster in the Northern sky, A2029, borders
the Bootes void and is located in a supercluster with four members,
SCL 154, in the filament between the Hercules and the Corona Borealis
superclusters (SCL 160 and 158).  Markevitch \etal (1998, hereafter
MFSV) describe this cluster as one of the most regular, well relaxed
X-ray cluster with a very strong cooling flow.

The third brightest X-ray cluster in the RBSC catalog is A401 which
forms a cluster pair with A399 (SCL 45).  Both of these clusters
contain a cD galaxy. MFSV suggest that these clusters may be in the
early stages of a collision.

Another isolated cluster of high X-ray luminosity, A478, shows 
evidence for a strong cooling flow (MFSV and White, Jones and Forman
1997). In clusters A478 and A2142 the Sunyaev-Zeldovich effect has been
measured (Myers \etal 1998).

One of the clusters of the highest X-ray luminosity in the DFJ sample
is A426, a cooling flow cluster (White, Jones and Forman 1997) in the
Perseus supercluster (SCL 40).

The brightest X-ray cluster in the sample by de Grandi \etal (1999),
A3266, is located in the outer region of the Horologium-Reticulum
supercluster (SCL 48), i.e. also in a relatively low-density
environment.  MFSV and Henriksen \etal (2000) show the possibility of
a merger event in this cluster.

The second brightest X-ray cluster in the sample by de Grandi \etal
(1999), A3186, is one of the most distant clusters in our sample lying
at a distance of about $350$ \Mpc\ in an area of a low-density
filament that surrounds distant voids in the Southern sky.  This cD
cluster shows evidence of a substructure and a small cooling flow
(Nesci and Norci 1997).  A3186 is of richness class $R = 1$, while all
other clusters of the highest X-ray luminosity mentioned here are of
richness class $R = 2$.

The third brightest cluster in de Grandi's sample is A3827, an
outlying member of the poor supercluster SCL~200.  X-ray emission of
this cluster is probably dominated by its central galaxy that shows
signs of merging of other galaxies in the cluster (Astronomy Picture
of the Day, August 31, 1998,
http://antwrp.gsfc.nasa.gov/apod/astropix.html).

\section{Discussion and conclusions}

We have studied the distribution of X-ray clusters with respect to the
supercluster-void network determined by Abell clusters, compiled a
list of X-ray clusters in superclusters and showed that both X-ray and
optical clusters delineate large-scale structure in a similar way.
X-ray clusters that do not belong to superclusters determined by Abell
clusters border the Southern and Northern Local supervoid or are
located in filaments between superclusters.  X-ray clusters are more
strongly clustered than optically selected clusters: the fraction of
X-ray clusters is higher in rich and very rich superclusters, and the
fraction of isolated X-ray clusters is lower than these fractions for
optically selected clusters.  These results indicate that the
structure of the Universe is traced in a similar way by both optical
and X-ray clusters up to redshifts of $z = 0.13$.  A similar
conclusion has been obtained by Borgani \& Guzzo (2001) based on the
comparison of the REFLEX cluster surveys with the Las Campanas galaxy
redshift survey (Shectman et al. 1996).

The rather regular placement of superclusters is noticeable in the
case of both X-ray clusters and Abell clusters, especially in the
Northern sky.  We shall discuss the presence of the regularity in the
distribution of X-ray clusters in more detail in Paper II. In
particular, we shall present evidence for a presence of a
characteristic scale of $120$\Mpc~ in the distribution of X-ray
clusters.

EETDA demonstrated that the fraction of X-ray clusters in
superclusters increases with supercluster richness (Table~4 in EETDA).
This result was based on the early catalogs of X-ray clusters
containing altogether 59 X-ray clusters in superclusters.  Our present
study confirms and even strengthens this early result.  The data in
Table~1 show that the fraction of X-ray clusters in the Abell
cluster-based superclusters increases with supercluster richness. In
several superclusters most members are X-ray sources. The presence of
X-ray emitting gas in a large fraction of clusters shows that
potential wells in clusters and superclusters of galaxies are rather
deep.

We did not detect a correlation between the X-ray luminosity of
clusters and their host supercluster richness, although clusters with
the highest X-ray luminosities are located in relatively poor
superclusters.

Loken \etal (1999) showed that massive cooling flow clusters are
located in high density regions. We find that from 26 clusters
analyzed in their study 24 belong to superclusters, and 12 of them to
very rich superclusters. Six clusters are members of the Hercules
supercluster.

Engels \etal (1999) found indications that X-ray selected AGNs may be
a part of the supercluster-void network described previously by
Einasto and co-workers (see references in the Introduction).  Our
results confirm this.  A number of AGNs from the RBS catalog are
located in superclusters of Abell clusters.  Several structures seen
in the distribution of X-ray selected AGN are also seen in our sample
(in the direction of the Pisces, the Ursa Majoris and the Coma
superclusters), although, in general, Engels \etal study more distant
objects beyond the borders of our sample.

Boughn (1999) demonstrated the presence of X-ray emission from the
Local supercluster as a possible evidence of hot diffuse gas in
superclusters.  Scharf \etal (2000) found an evidence for X-ray
emission from a distant large scale filament of galaxies.  In the
Shapley supercluster X-ray emission has been detected in the filaments
between supercluster member clusters (Bardelli \etal 1999, Kull and
B\"ohringer 1999).  This indicates that the whole central part of the
supercluster is a physical entity forming a deep potential well. 

These findings give additional evidence that superclusters are not
random associations of clusters but form real physical systems --
large-scale high-density regions of the matter distribution forming
extended potential wells in the distribution of matter.  Both optical
and X-ray clusters are parts of the same supercluster-void network
that we see in the distribution of Abell clusters of galaxies.  Our
results suggest that optically and X-ray selected cluster samples can
be used to find large-scale high-density regions in the Universe.
Samples detected optically and in X-rays are different in many
details, but are common in one important aspect -- both indicate the
skeleton of the supercluster-void network in a rather similar way.

Main results of our study of the clustering properties of X-ray
clusters are:

1) We present an updated catalog of superclusters of Abell
clusters and  a list of X-ray clusters in superclusters.  

2) Optical and X-ray clusters trace the supercluster-void network in
a similar way.

3) The fraction of X-ray clusters in superclusters increases with
the supercluster richness suggesting that superclusters are real
physical systems.

4) Cluster X-ray luminosity is not correlated with their host
supercluster richness, although the most luminous X-ray clusters are
located in relatively low density environments.

\section*{Appendix A: A catalog of superclusters of Abell clusters}

Here we present a new supercluster catalog based on the Abell cluster sample
(A1) used in this paper.

The catalog of superclusters of Abell clusters is based on a cluster
sample which contains all superclusters of richness class $N_{CL} \geq
2$.  Table~A1 contains the following entries: $No$ is the
identification number.  The supercluster should be referred to as
``SCL~nnn'' with nnn being the running number $No$.  As mentioned in
the text, an index "c" in the first column indicates a supercluster
candidate, i.e. a supercluster that is not present in the test
catalog determined by clusters of measured redshifts only.

$N_{CL}$ is the number of member clusters in the supercluster;
$\alpha_C$ and $\delta_C$ are coordinates of the center of the
supercluster (equinox 1950.0), derived from coordinates of individual
clusters; $D_C$ is the distance of the center from us; it follows the
list of Abell clusters which are members of the supercluster. An index
"e" after the Abell cluster number in the column 6 shows that this
cluster has only an estimated distance.  In the last column we list a
commonly used name of the supercluster, which in most cases is based
on constellation names.  To avoid confusion, we use the same numbers
as in our previous version of the catalog (E97d); and add new
numbers (221 and above) for superclusters described in this catalog
for the first time. Superclusters are sorted by $\alpha_{C}$.

\section*{Appendix B: X-ray clusters in superclusters}

In Table B1 we present  data on  X-ray clusters in superclusters, 
while  Table B2 lists additional systems of X-ray clusters. Columns 
for both tables are as  follows: 
 
(1) identification number of the supercluster in the catalog; 
subscript $C$ means supercluster candidate; 
 
(2) Abell numbers of all clusters in the supercluster, 
according to Table A1; 
 
(3) , (4) and (5) -- center coordinates for the  supercluster ($\alpha$,
$\delta$ and distance to the supercluster center);

(6): Catalog numbers of X-ray clusters in the supercluster. We use
Abell - ACO catalog numbers for clusters identified in this
catalog.  Cluster numbers without subscript are from RBSC catalog;
index $G$ means clusters from de Grandi \etal (1999) catalog
only, index $D$ means clusters from the DFJ catalog only, index
$B$ -- clusters from the BCS catalog only.

Double subscripts refer to non-Abell clusters.  Index $RR$ means
clusters number from RBS catalog; index $BB$ -- cluster
number from BCS catalog; index $GG$ -- cluster number from
the catalog by de Grandi \etal (1999).

In Table B2 clusters without subscripts refer to Abell clusters
that are not listed in the X-ray cluster catalogs used in the
present study.

(7): identification of supercluster.

\acknowledgments We thank G\"unther Hasinger for providing us with a
draft version of the RBS catalog and discussion of preliminary results
of the study. We thank Enn Saar and Alexei Starobinsky for stimulating
discussion.  This work was supported by Estonian Science Foundation
grant 2625. JE thanks Astrophysical Institute Potsdam for hospitality
where part of this study was performed.  HA thanks CONACyT for
financial support under grant 27602-E.

\vfill\eject

{\tiny
\vskip-2cm
\begin{table*}
\tablenum{A1}
\begin{center}
\caption{The list of superclusters}
\begin{tabular}{rrrrrrrrrrrrrrrr}
\\
\hline
\\
(1)& (2) & (3) & (4) &(5) & &&&&(6)&&&&& & (7)  \\
\\
$No$ & $N_{CL}$ & $\alpha_C$ & $\delta_C$ & $D_C$ 
&&\multispan4 Abell-ACO No.&&&& &  \\
  & &  &  &\multispan1$ h^{-1} Mpc$ &&&&&& & &&&& \\

\\
\hline
\\
\end{tabular}
\end{center}
\end{table*}

\begin{table*}
\tablenum{A1}
\vskip-2cm
\begin{center}
\caption{...continued}
\begin{tabular}{rrrrrrrrrrrrrrrr}
\\
\hline
\\
(1)& (2) & (3) & (4) &(5) & &&&&(6)&&&&&\multispan2 (7)  \\
\\
$No$ & $N_{CL}$ & $\alpha_C$ & $\delta_C$ & $D_C$ 
&&\multispan4 Abell-ACO No.&&&& \multispan2    \\
  & &  &  &\multispan1$ h^{-1} Mpc$ &&&&&& & &&&& \\

\\
\hline
\\
\end{tabular}
\end{center}
\end{table*}

\begin{table*}
\tablenum{A1}
\vskip-2cm
\begin{center}
\caption{...continued}
\begin{tabular}{rrrrrrrrrrrrrrrr}
\\
\hline
\\
(1)& (2) & (3) & (4) &(5) & &&&&(6)&&&&&\multispan2 (7)  \\
\\
$No$ & $N_{CL}$ & $\alpha_C$ & $\delta_C$ & $D_C$ 
&&\multispan4 Abell-ACO No.&&&& \multispan2    \\
  & &  &  &\multispan1$ h^{-1} Mpc$ &&&&&& & &&&& \\

\\
\hline
\\
\end{tabular}
\end{center}
\end{table*}

\begin{table*}
\tablenum{A1}
\vskip-2cm
\begin{center}
\caption{...continued}
\begin{tabular}{rrrrrrrrrrrrrrrr}
\\
\hline
\\
(1)& (2) & (3) & (4) &(5) & &&&&(6)&&&&&\multispan2 (7)  \\
\\
$No$ & $N_{CL}$ & $\alpha_C$ & $\delta_C$ & $D_C$ 
&&\multispan4 Abell-ACO No.&&&& \multispan2    \\
  & &  &  &\multispan1$ h^{-1} Mpc$ &&&&&& & &&&& \\

\\
\hline
\\
\end{tabular}
\end{center}
\end{table*}
}

\vfill\eject

{\tiny
\vskip-1cm
\begin{table*}
\tablenum{B1}
\begin{center}
\caption{The list of X-ray clusters in superclusters}
\begin{tabular}{rrrrrrrrrr}
\\
\hline
\\
(1)& (2) & (3) & (4) &(5) &&&(6)&& (7)  \\
\\
$No$ & $N_{CL}$ & $\alpha_C$ & $\delta_C$ & $D_C$ &
\multispan4 Cluster  No. \\
  & &  &  &\multispan1$ h^{-1} Mpc$ &&&&& \\

\\
\hline
\\
\end{tabular}
\end{center}
\end{table*}
}
\vfill\eject

{
\vskip-1cm
\begin{table*}
\tablenum{B2}
\begin{center}
\caption{The list of additional superclusters of non-Abell X-ray clusters}
\begin{tabular}{rrrrrrrrr}
\\
\hline
\\
(1)& (2) & (3) & (4) &(5) &&&(6)&\\
\\
$No$ & $N_{CL}$ & $\alpha_C$ & $\delta_C$ & $D_C$ &
\multispan3 Cluster  No. \\
  & &  &  &\multispan1$ h^{-1} Mpc$ &&&& \\

\\
\hline
\\

     1 &     2 &     5.2 &    28.9 &    264 &
    21 &     4\rlap{BB} &
  &
 
  \\ 
     2 &     2 &    11.6 &    25.0 &    232 &
   104\rlap{RR} &
   104 &  &
  
  \\ 
     3 &     2 &    27.2 &    -4.6 &    117 &
   295 &
    23\rlap{GG} &  &
  
  \\ 
     4 &     4 &    32.1 &     2.7 &     59 &
   194 &    
   400 &
   199\rlap{RR} &
26\rlap{GG} 
  \\ 
     5 &     2 &    32.4 &    31.3 &    103 &
   260 &    25\rlap{BB} &
  &
  
  \\ 
     6 &     2 &    34.8 &   -27.6 &    169 &
  2992 &   317\rlap{RR} &
  &
  
  \\ 
     7 &     3 &    51.4 &    14.2 &     93 &
   397 &
   456\rlap{RR} &
   461\rlap{RR} &
  
  \\ 
     8 &     2 &    55.6 &   -54.4 &    127 &
  3144 &   485\rlap{RR} &
  &
  
  \\ 
    9 &     3 &    68.8 &   -12.1 &    104 &
   496 &    44\rlap{GG} &
   540\rlap{RR} &
  
   \\ 
    10 &     2 &    94.7 &   -64.8 &     73 &
  3389 &   62\rlap{GG} &
  &
  
  \\ 
    11 &     2 &   114.0 &    52.8 &    186 &
   595 &    44\rlap{BB} &
  &
  
  \\ 
    12 &     5 &   114.9 &    54.3 &     81 &
   569 &
   576 &
   634 &    47\rlap{BB} 
\\
&&&&&
    48\rlap{BB} &
  &
  &
  
  \\ 
    13 &     2 &   215.1 &    48.5 &    201 &
  1904 &  1380\rlap{RR} &
  &
  
  \\ 
    14 &     2 &   222.9 &    22.3 &    271 &
  2021 &   130\rlap{BB} &
  &
  
  \\ 
    15 &     2 &   305.9 &   -20.2 &    160 &
   2324 &   76\rlap{GG} &
  &
  
  \\ 
    16 &     2 &   315.5 &   -52.2 &    136 &
  3716 &  1719\rlap{RR} &
  &
  
  \\ 
    17 &     2 &   327.4 &   -44.6 &    173 &
   3809 &   91\rlap{GG} &
  &
  
  \\ 
    18 &     2 &   350.9 &   -40.5 &    161 &
   4008 &  122\rlap{GG} &
  &
  
  \\ 
    19 &     2 &   356.1 &    -3.8 &    221 &
   2656 & 2042\rlap{RR} &
  &
  
  \\

\label{tab:sc2}
\end{tabular}
\end{center}
\end{table*}
}
\vfill\eject

\end{document}